\documentclass{aastex62}

\usepackage{amsmath}	

\definecolor{awesome}{rgb}{1.0, 0.13, 0.32}
\definecolor{darkcyan}{rgb}{0.0, 0.55, 0.55}
\definecolor{darkgreen}{rgb}{0.13, 0.55, 0.13}

\graphicspath{{./}{figures/}}

\accepted{\today}
\submitjournal{ApJ}

\shorttitle{Impact of ionospheric activity on EoR Power Spectrum with the MWA}
\shortauthors{Trott et al.}

\begin{document}

\title{Assessment of ionospheric activity tolerances for Epoch of Reionisation science with the Murchison Widefield Array}

\correspondingauthor{C.~M. Trott}
\email{cathryn.trott@.curtin.edu.au}

\author[0000-0001-6324-1766]{Cathryn M. Trott}
\affil{International Centre for Radio Astronomy Research - Curtin University \\
1 Turner Avenue, Bentley
WA 6102, Australia}
\affiliation{ARC Centre of Excellence for All Sky Astrophysics in 3 Dimensions (ASTRO 3D)}

\author{C.~H. Jordan}
\affil{International Centre for Radio Astronomy Research - Curtin University \\
1 Turner Avenue, Bentley
WA 6102, Australia}
\affiliation{ARC Centre of Excellence for All Sky Astrophysics in 3 Dimensions (ASTRO 3D)}

\author{S.~G. Murray}
\affil{International Centre for Radio Astronomy Research - Curtin University \\
1 Turner Avenue, Bentley
WA 6102, Australia}
\affiliation{ARC Centre of Excellence for All Sky Astrophysics in 3 Dimensions (ASTRO 3D)}

\author{B. Pindor}
\affil{The University of Melbourne, School of Physics, Parkville, VIC 3010, Australia}
\affiliation{ARC Centre of Excellence for All Sky Astrophysics in 3 Dimensions (ASTRO 3D)}

\author{D.~A. Mitchell}
\affil{CSIRO Astronomy and Space Science (CASS), PO Box 76, Epping, NSW 1710, Australia}
\affiliation{ARC Centre of Excellence for All Sky Astrophysics in 3 Dimensions (ASTRO 3D)}

\author{R.~B. Wayth}
\affil{International Centre for Radio Astronomy Research - Curtin University \\
1 Turner Avenue, Bentley
WA 6102, Australia}
\affiliation{ARC Centre of Excellence for All Sky Astrophysics in 3 Dimensions (ASTRO 3D)}

\author{J.~Line}
\affil{The University of Melbourne, School of Physics, Parkville, VIC 3010, Australia}
\affiliation{ARC Centre of Excellence for All Sky Astrophysics in 3 Dimensions (ASTRO 3D)}

\author{B.~McKinley}
\affil{International Centre for Radio Astronomy Research - Curtin University \\
1 Turner Avenue, Bentley
WA 6102, Australia}
\affiliation{ARC Centre of Excellence for All Sky Astrophysics in 3 Dimensions (ASTRO 3D)}

\author{A.~Beardsley}
\affil{Arizona State University, USA}

\author{J.~Bowman}
\affil{Arizona State University, USA}

\author{F.~Briggs}
\affil{The Australian National University, Australia}

\author{B.~J.~Hazelton}
\affil{University of Washington, Department of Physics, Seattle, WA 98195, USA}

\author{J.~Hewitt}
\affil{Massachusetts Institute of Technology, Cambridge, MA, USA}

\author{D.~Jacobs}
\affil{Arizona State University, USA}

\author{M.~F.~Morales}
\affil{University of Washington, Department of Physics, Seattle, WA 98195, USA}

\author{J.~C.~Pober}
\affil{Brown University, Department of Physics, Providence, RI 02912, USA}

\author{S.~Sethi}
\affil{Raman Research Institute, Bangalore, India}

\author{U.~Shankar}
\affil{Raman Research Institute, Bangalore, India}

\author{R.~Subrahmanyan}
\affil{Raman Research Institute, Bangalore, India}

\author{M.~Tegmark}
\affil{MIT Kavli Institute, Massachusetts Institute of Technology, Cambridge, MA, USA}

\author{S.~J.~Tingay}
\affil{International Centre for Radio Astronomy Research - Curtin University \\
1 Turner Avenue, Bentley
WA 6102, Australia}
\affiliation{ARC Centre of Excellence for All Sky Astrophysics in 3 Dimensions (ASTRO 3D)}

\author{R.~L.~Webster}
\affil{The University of Melbourne, School of Physics, Parkville, VIC 3010, Australia}
\affiliation{ARC Centre of Excellence for All Sky Astrophysics in 3 Dimensions (ASTRO 3D)}

\author{J.~S.~B.~Wyithe}
\affil{The University of Melbourne, School of Physics, Parkville, VIC 3010, Australia}
\affiliation{ARC Centre of Excellence for All Sky Astrophysics in 3 Dimensions (ASTRO 3D)}

\begin{abstract}
Structure imprinted in foreground extragalactic point sources by ionospheric refraction has the potential to contaminate Epoch of Reionisation (EoR) power spectra of the 21~cm emission line of neutral hydrogen. The alteration of the spatial and spectral structure of foreground measurements due to total electron content (TEC) gradients in the ionosphere create a departure from the expected sky signal. We present a general framework for understanding the signatures of ionospheric behaviour in the two-dimensional (2D) neutral hydrogen power spectrum measured by a low-frequency radio interferometer. Two primary classes of ionospheric behaviour are considered, corresponding to dominant modes observed in Murchison Widefield Array (MWA) EoR data; namely, anisotropic structured wave behaviour, and isotropic turbulence. Analytic predictions for power spectrum bias due to this contamination are computed, and compared with simulations.
We then apply the ionospheric metric described in \citet{jordan16} to study the impact of ionospheric structure on MWA data, by dividing MWA EoR datasets into classes with good and poor ionospheric conditions, using sets of matched 30-minute observations from 2014 September. The results are compared with the analytic and simulated predictions, demonstrating the observed bias in the power spectrum when the ionosphere is active (displays coherent structures or isotropic turbulence). The analysis demonstrates that unless ionospheric activity can be quantified and corrected, active data should not be included in EoR analysis in order to avoid systematic biases in cosmological power spectra. When data are corrected with a model formed from the calibration information, bias reduces below the expected 21~cm signal level. Data are considered `quiet' when the median measured source position offsets are less than 10--15~arcseconds.
\end{abstract}




\section{Introduction}
Detection and characterisation of the neutral hydrogen signal from the early Universe provides one of the best probes for exploring cosmology and source astrophysics at these epochs \citep{furlanetto06}. As an emission-line signal emitted by the neutral intergalactic medium (IGM), study of its spatial and redshift structure, and amplitude, encodes information about the growth of structure, and evolution of the first ionising sources of radiation. It therefore offers the potential to connect CMB observations from the last scattering surface with low- and intermediate-redshift observations of AGN and galaxies, to obtain a complete evolutionary history of the Universe.

Measurements of the integrated optical depth to electrons \citep{planck15} and of the Ly-$\alpha$ Forest along sight lines to high redshift quasars \citep{fan06,mortlock16}, constrain the redshift range for the EoR, where the averaged neutral fraction of the IGM reduces from 95\% to 5\%, between $z=[5.5-10]$, placing observations of the rest frame 1420~MHz neutral hydrogen emission line within reach of low-frequency instruments ($\nu=[130-220]$~MHz). The current low-frequency telescopes that are attempting to detect the EoR signal at lower redshift include the Murchison Widefield Array, MWA{\footnote[1]{http://www.mwatelescope.org}} \citep{bowman13_mwascience,tingay13_mwasystem,jacobs16}; the Precision Array for Probing the Epoch of Reionization, PAPER{\footnote[2]{http://eor.berkeley.edu}} \citep{parsons10}; the Low Frequency Array, LOFAR{\footnote[3]{http://www.lofar.org}} \citep{vanhaarlem13,patil16}; the Long Wavelength Array, LWA{\footnote[4]{http://lwa.unm.edu}} \citep{ellingson09}. These will soon be complemented by second-generation experiments; HERA \citep{deboer16} and Square Kilometre Array Low Telescope (SKA-Low) \citep{koopmans15}. While current experiments aim to detect the EoR signal, and provide some broad evolutionary parameters, future experiments aim to explore this era, yielding a wealth of cosmological and astrophysical information.

Despite the substantial current efforts to detect the EoR signal from neutral hydrogen, a spatial fluctuation detection has yet eluded the community \citep[unlike the global signal, potentially detected in the Cosmic Dawn by][]{bowman18}. While the signal weakness (tens of mK brightness temperature fluctuations) relative to measurement noise requires 100s--1000s~hours of observations, the principal barriers are the complexity of the instruments and bright foreground (extragalactic and Galactic) emission. These foreground contaminants, observed through complex and chromatic radio interferometers, yield contamination in the signal across the parameter space of interest (angular and line-of-sight spatial scales). Careful instrument calibration and foreground treatment is crucial for accessing the weak, high-redshift signal \citep{trottwayth2016,barry16,patil16}. Any unmodelled structure to the foregrounds therefore presents a challenge, altering the signal processing that needs to be undertaken to perform the experiment, and having the potential to imprint residual structure that mimics the cosmological signal. One such source of unmodelled structure is refraction of the incoming foreground wavefronts by plasma non-uniformity in the Earth's upper atmosphere, principally the ionosphere \citep{loi16,mevius16,jordan16,degasperin18}. Gradients in the plasma across the pierce points to extragalactic sources yield source position offsets, while curvature induces flux density changes (scintillation) as the wavefront acquires concavity or convexity \citep{vedantham16}. \citet{tasse13}, \citet{prasad14}, \citet{martividal10}, \citet{wijnholds10}, \citet{cornwell16} and \citet{hurleywalker18} study the impact of the ionosphere on calibration and offer algorithms for correcting the direction-dependent effects. Most recently, \citet{rioja18} have suggested the LEAP algorithm to provide post-imaging corrections to the data, with application to the extended array configuration of MWA Phase II and SKA. Long baseline arrays (e.g., LOFAR) have observed source decoherence \citep{mevius16}, and these have been the basis for much of the early work in this area \citep{vandertol07}. While \citet{vedantham16} discussed the implications for EoR of scintillation noise on interferometric visibilities, we study the impact of residual noise power bias from refractive positional offsets of foreground point sources, which hinder our view of the weak cosmological signal.

The paper is structured as follows. In \S \ref{section:methods}, we present an overview of the MWA EoR experiment, the data collected for this work, and the quantitative analysis described in \citet{jordan16}. The analytic models for the ionosphere are then described in \S \ref{section:models}, and \S \ref{section:results} with details of the simulations used to support the analysis. Finally, in \S \ref{section:tolerances}, we apply the models to the distribution of ionospheric conditions observed at the MRO and characterised by \citet{jordan16}, and predict the impact on MWA EoR experiments. Throughout, vectors are denoted with an over-arrow, matrices are upper case and in script, and estimates of a quantity have an over-hat. Fourier Transforms are denoted with a script, $\mathcal{F}$.

\section{Methods}\label{section:methods}
\subsection{The MWA EoR Experiment}
The MWA \citep{tingay13_mwasystem} is a low-frequency aperture array interferometer, located in the Western Australian desert on the same site as the future SKA-Low, and is an SKA precursor instrument. Its science goals are shared with many other low-frequency instruments; namely, early Universe studies through redshifted neutral hydrogen, pulsar searches and studies, fast and slow transients, and low-frequency sky catalogues \citep{bowman13_mwascience,wayth15,hurleywalker16}. In Phase I (2013--2016), relevant for this work, it comprised 128 tiles of 16 dual-polarisation crossed dipole antennas, spread over a diameter of $\sim$3~km.

The MWA EoR experiment \citep{jacobs16} observes two primary fields (EOR0 and EOR1) in cold parts of the radio sky, with a total observing time exceeding 1500~hours, with the principal goal of statistical detection of the EoR, and potential for estimation of the slope and amplitude of the spherically-averaged power spectrum. It operates over two primary frequency bands: 138--167~MHz and 167--197~MHz.

\subsection{Ionospheric conditions at the MRO}
\citet{beardsley16} have published the deepest upper limits on the EoR power spectrum in the redshift range $z=6.5-7.5$ for the MWA EoR experiments, with inclusion of 32-hours of high-quality EoR0 data in the high-band (167--197~MHz), derived from a raw sample of 86-hours. Much of the data that were unused in the analysis comes from observations with poor calibration, attributable to periods of substantial ionospheric activity \citep{jordan16}. At small $k$, they found that the data were systematics-dominated, and attributed this excess power to foreground contamination. These published data, which included some remaining ionospheric activity (identified more recently with new techniques) were processed through the two independent analysis pipelines that are described in \citet{jacobs16}. This dual processing ensured consistency and reliability of results, and the consequent processing through the MWA Real Time calibration System \citep[RTS,][]{mitchell08} resulted in calibration log files being available for the final dataset. \citet{jordan16} analysed these calibration files as a function of time and sky location, deriving ionospheric characteristics across $\sim$1000 point sources each 8~seconds. The analysis of these data resulted in the development of a metric to quantitatively describe the degree of ionospheric activity, and identification and classification of four distinct ionospheric types. The metric has two independent components, which capture the two primary classes of activity observed: isotropic turbulent-like increases in the source position variations compared with thermal noise, and anisotropic structured waves of plasma density with a spatially periodic structure \citep[cf,][who discovered these density ducts in MWA datasets]{loi16}. The relevant components of the metric for these quantify the median absolute source offset, relative to a catalogue position, and the anisotropy of position offset vectors (quantified via a Principal Component Analysis of the ensemble source vectors), respectively. See \citet{jordan16} for relevant details.

In this work, we combine these two structural models to form a generic and tunable model for ionospheric structure, akin to that developed by \citet{intema_thesis09,intema14} and \citet{vandertol07}, which also summed a turbulent and trigonometric component. The ionospheric phase screen derivative at position  $l,m$ and frequency $\nu$, is parametrised as:
\begin{equation}
\phi(l,m,\nu;k_l,k_m,r_0,n) = A(\nu)\phi_{\rm ani}(l,m,\nu;k_l,k_m) + B(\nu)\phi_{\rm turb}(l,m,\nu;r_0,n),
\end{equation}
where $k_l,k_m$ parametrise the spatial wavenumbers of the anisotropic component, and $r_0, n$ parametrise the characteristic scale and power-law exponent of the turbulent component. For the purposes of demonstrating the four classes of ionospheric conditions, the overall frequency-dependent amplitude of these components is controlled by $A(\nu)$ and $B(\nu)$, with a frequency-squared dependence demonstrated in the data \citep{jordan16}.

The anisotropic component is modelled explicitly by the observed \textit{variation} of source number density compared with a stochastic (unclustered) field, such that:
\begin{equation}
\Delta\vec{l} (l,m,\nu;k_l) \propto \sin{(2\pi k_l l)},
\end{equation}
where the shift in the apparent source position, $\Delta\vec{l}$ is related to the gradient of the TEC screen (divergence of the source number density):
\begin{equation}
\nabla\phi_{\rm ani}(l,m,\nu;k_l) \propto \Delta\vec{l}.
\end{equation}
The fields used by the MWA EoR experiment do not demonstrate any clustering for sources with $S>100$~mJy (the confusion limit).

The turbulent component exhibits power-law behaviour in Fourier space, and is defined by its phase structure function (which is equivalent to the sum of the two-point correlation function and the variance -- correlation at zero lag),
\begin{equation}
D_{\phi\phi}(r) = \langle [\phi_{\rm turb}(r+p)-\phi_{\rm turb}(p)]^2\rangle_p = \left(\frac{r}{r_0}\right)^{n},
\end{equation}
where $p$ characterises the spatial separation of points on the celestial sphere.

The four primary classes, and their relative occupancy in the MWA EoR data, identified by \citet{jordan16} are:
\begin{itemize}
\item Type I: quiet ionosphere; $A, B$ small (74\%);
\item Type II: turbulent ionosphere; $A$ small, $B$ large (15\%);
\item Type III: structured, anisotropic ionosphere; A moderate, B small (3\%);
\item Type IV: structured and turbulent ionosphere; A, B large (8\%).
\end{itemize}
Thus, the bulk of the data were acquired during quiet or isotropic turbulent modes. These data are all using zenith-pointed primary beams, where the phase centre HA$<$1~hour. We now employ these types to form analytic models for EoR data, and propagate into the 2D power spectrum. We then use the relative occupancy of data of each type to predict the impact on current and future EoR experiments at the MRO. For active modes, we assess the two qualitatively-different modes in these active types: anisotropic, structured modes, and isotropic turbulence.

\subsection{Power spectrum}
We aim to derive the expected power spectrum of the extragalactic point sources due to ionospheric disturbance of the wavefront. The power spectrum of neutral hydrogen brightness temperature fluctuations is defined as:
\begin{equation}
P_{21}(k=|\vec{k}|) = \frac{\delta(\vec{k}-\vec{k}^\prime)}{\Omega} \,\, \langle \tilde{T}(\vec{k})\tilde{T}(\vec{k}^\prime)\rangle  \,\,{\rm mK}^2 h^{-3}{\rm Mpc}^3,
\end{equation}
where $\Omega$ denotes the observation volume and $T$ is the 21-cm brightness temperature field, and the delta function encodes the expectation of signal isotropy. In the flat-sky approximation, where the Fourier-transform of the sky brightness can be equated with the measured interferometric visibilities, we can use the distribution of visibilities to estimate the power spectrum of the point source foregrounds:
\begin{equation}
P(k) \propto \langle V^\ast(k) V(k) \rangle = {\rm Var}(V(k)) + \langle V(k) \rangle\langle V^\ast(k) \rangle,
\end{equation}
where we have used the identity for the variance of a quantity in terms of its first- and second-moments. To understand the effect of the ionosphere on the power spectrum, we therefore only are required to determine the expected value and variance of the measured visibilities under active and quiet ionospheric conditions, where the former is studied for both turbulent and structured activity.

\section{Ionospheric models}\label{section:models}
Motivated by the knowledge that foreground power and structure will remain a primary impediment to detecting and characterising the EoR cosmological signal, we are interested in understanding the impact of ionospheric activity on residual foreground signals. For this work, we employ a frequency-dependent model of point source foregrounds (radio and star-forming galaxies at $z<6$, but primarily below $z=2$), to represent the residual sky model.

\subsection{Foreground point source models}\label{section:analytic_model}
{\bf Turbulent component:} The isotropic component of the model is used to represent a Kolmogorov turbulence in the plasma. We follow the formalism of \citet{murray16}, who developed a generic model for the statistical signature of clustering of extragalactic point sources in the 2D power spectrum. In addition to the intrinsic cosmological clustering of AGN and star forming galaxies, they extend the formalism to derive the power spectrum of apparent source overdensities due to a Kolmogorov turbulence in the ionosphere. This derivation is based on the weak-lensing formalism of the spatial distortion of a field according to a Total Electron Content (TEC) gradient. The full derivation is presented in the Appendix, and the main results reproduced here.

We denote the (scalar) TEC field by $\psi(\vec{l})$ and note that in the weak-field regime with which we are here concerned, the first-order effect of the ionosphere is to impose a shift on the observed angular position of incoming sight-lines. 
This shift can be written
\begin{equation}
\label{eq:tec_shift}
    \Delta \vec{l} = K(\nu) {\nabla} \psi(\vec{l}),
\end{equation}
where $K \propto \nu^{-2}$ is a scaling constant.
The observed power spectrum \citep[cf.][Eq. 4.12]{Lewis2008} is:
\begin{equation}
    \label{eq:power_observed}
  {P}_{\rm turb} \approx (1 - K(\nu) u^2 R^\psi) P_0(u) + K(\nu) P_u^\psi \star P_0(u),
\end{equation}
with $\star$ indicating convolution, $u$ encoding spatial wavenumber, and
\begin{equation}
R^\psi = \pi \int u^3 P_u^\psi du.
\end{equation}
The second term in Equation \ref{eq:power_observed} is sub-dominant, and we ignore it herein.
We suppose that the power-spectrum of $\psi$ is Kolmogorov-like, but has some scale above which the power is truncated. In practice, setting this scale above the resolution limit of an instrument serves to eliminate sensitivity to this choice. 
Thus we consider 
\begin{equation} 
P^{\psi}_u = \begin{cases}
\left(\frac{u}{u_0}\right)^\kappa & u \leq u_{\rm max} \\
0 & u>u_{\rm max},
\end{cases}
\end{equation}
where $u_0$ sets the characteristic scale of the turbulence.
In this case, we have 
\begin{equation}
R^\psi = \frac{u_{\rm max}^{4+\kappa}}{(4+\kappa)u_0^\kappa}.
\end{equation}
Thus, the expected angular power spectrum due to Kolmogorov-like turbulence modifies the unperturbed power according to:
\begin{equation}
{P}_{\rm turb} = \left( 1 - K(\nu) u^2 \frac{u_{\rm max}^{4+\kappa}}{(4+\kappa)u_0^\kappa} \right) P_u^S.
\end{equation}

The last step is to include the line-of-sight contribution, where the amplitude of the turbulence follows a frequency-squared dependence, $K_u \propto \nu^{-2}$. For the prediction, we are assuming a perfect instrument, whereby the Fourier plane is completely filled and the `wedge'-like signature of smooth foregrounds (due to incomplete interferometric sampling) is replaced by a `brick'-like feature where power occupies the $\eta=0$ mode and small $\eta$ modes.
If the ionospheric dependence on frequency and spatial scale are assumed to be independent, we can approximately decouple the angular and line-of-sight Fourier Transforms, yielding:
\begin{equation}
{P}_{\rm turb} = P_{u\eta}^S W_0(\eta) - K_0 W_{2}(\eta) u^2 \frac{u_{\rm max}^{4+\kappa}}{(4+\kappa)u_0^\kappa} P_{u\eta}^S,
\label{eqn:turb}
\end{equation}
where,
\begin{equation}
W_{2}(\eta) = |\mathcal{F}(W(\nu)\nu^{\gamma-2})|^2,
\end{equation}
and,
\begin{equation}
W_0(\eta) = |\mathcal{F}(W(\nu)\nu^{\gamma})|^2,
\end{equation}
are the Fourier Transforms of the window function multiplied with the decaying spectral behaviour of the ionosphere and foreground sources (index $\gamma\sim -0.7$), and the bandwidth window function multiplied with the spectral behaviour of the foregrounds alone, respectively.
This term encodes the difference in line-of-sight spatial mode behaviour for differing power-law dependencies on the frequency. In practice, for observational volumes with sufficient fractional bandwidth, the additional frequency-dependence of the ionosphere will yield an additional leakage of foreground power from the brick into the EoR Window. This effect is shown in \S \ref{section:tolerances}.
\vspace{1cm}

{\bf Structured component:} 
In \citet{trottchips2016}, we present a framework for representing the statistical 2D power spectral signature of unclustered, Poisson-distributed point-source foregrounds within an interferometric instrument with a frequency-dependent primary beam and chromatic Fourier sampling. To extend this simple model, where the sources are assumed to be distributed randomly on the sky, to one where sources are apparently clustered into periodic anisotropic tubes via ionospheric refraction, a term with sky position dependence is added to the source number density model, characterised by amplitude $\epsilon$, relative to the intrinsic case. Without loss of generality for a generic instrument, we consider the ionospheric perturbations to be aligned with $y$-axis, removing functional dependence on the $m$-component (North-South direction cosine):
\begin{equation}
\frac{dN}{dS}(l,m;\nu) = \alpha \left(1  + \epsilon(\nu) \sin{[2\pi  k_l l]}\right)\left( \frac{S_{\rm Jy}}{S_0}\right)^{-\beta}\,\, {\rm Jy^{-1} sr^{-1}}.
\label{eqn:anis_model}
\end{equation}
where
\begin{equation}
\epsilon(\nu) = \epsilon_0 \left( \frac{\nu}{\nu_0} \right)^{-2}.
\end{equation}
We assume a two-dimensional frequency-dependent primary beam with a Gaussian-shape. This is broadly representative of the beam shape for a square dipole-based tile (such as the MWA's 4x4 dipole arrangement), allows decoupling of the $l$ and $m$ planes, and has an analytic Fourier Transform:
\begin{equation}
B(l,n;\nu) = \exp{\left(-(l^2+m^2)/2\sigma_{\nu}^2 \right)}.
\end{equation}
Propagation of the source model through the frequency-dependent instrument model (primary beam, with characteristic size $\sigma_\nu$) yields an expression for the variance and expected value at angular Fourier modes $(u,v)$ and frequency $\nu$, given by:
\begin{eqnarray}
{\rm Var}(V(u,v)) &=& \frac{\alpha S_{\rm max}^{3-\beta}}{3-\beta} \pi\sigma_{\nu}^2 \left( 1 + \frac{\epsilon(\nu)^2}{2} - \frac{\epsilon(\nu)^2}{2}\exp{(-4\pi^2k_l^2\sigma_{\nu}^2)} \right)  \label{varsrc} \\
\langle V(u,v) \rangle &=& \frac{\alpha S_{\rm max}^{2-\beta}}{2-\beta} 2\pi\sigma_{\nu}^2 \exp{(-2\pi^2v^2\sigma_\nu^2)}\, {G},
\end{eqnarray}
where
\begin{eqnarray}
{G} &=& \exp{(-2\pi^2u^2\sigma_\nu^2)} \\\nonumber
&+& \frac{\epsilon(\nu)}{2} \left(\exp{(-2\pi^2\sigma^2(u-k_l)^2)} - \exp{(-2\pi^2\sigma^2(u+k_l)^2)} \right).
\end{eqnarray}
These expressions can then be combined to form the expected power spectrum due to extragalactic point source foregrounds distorted coherently by the ionosphere. The equivalent quiet ionosphere expressions can be recovered by setting $\epsilon=0$.

As with the turbulent component, the frequency-dependence of the anisotropic component can be approximated by applying the Fourier Transform of power-laws with different indices to the quiet and active components. If the point source foregrounds are considered to be spectrally-flat (a reasonable assumption over a small fractional bandwidth compared with the inverse-square dependence of the ionospheric component), then the spectral channels are completely correlated, and the $\eta=0$ term contains all of the power (up to spectral leakage due to a finite bandwidth, which causes some spillage of power into small $\eta$ modes, yielding the brick-like feature)\footnote{Expanding the variance and squared expectation value, and propagating the variance through the Fourier Transform, we obtain:
\begin{eqnarray}
&&\langle V(k,\eta)^\ast V(k,\eta)\rangle = \mathcal{F}^\dagger\langle V(k,\nu)^\ast V(k,\nu)\rangle\mathcal{F}\\
&=& \mathcal{F}^\dagger {\rm Var}(V_i + V_o) \mathcal{F} + \langle \mathcal{F}V_i + \mathcal{F}V_o\rangle^2 \nonumber \\
&=& \mathcal{F}^\dagger {\rm Var}(V_i) \mathcal{F} + \mathcal{F}^\dagger {\rm Var}(V_o) \mathcal{F} + \langle \mathcal{F}V_i \rangle^2 + \langle \mathcal{F}V_o \rangle^2 + 2\langle \mathcal{F}V_i \rangle \langle \mathcal{F}V_o \rangle \nonumber \\
&=& W_0{\rm Var}(V_o(k,\eta_0) + W_2 V_i(k,\eta_0)) + \langle \sqrt{W_0}V_o(k,\eta_0) + \sqrt{W_2}V_i(k,\eta_0) \rangle^2.\nonumber
\end{eqnarray}
}:
\begin{align}
\label{eqn:anis}
&\langle V(k,\eta)^\ast V(k,\eta)\rangle = \\\nonumber
&W_0{\rm Var}(V_o(k,\eta_0) + W_2 V_i(k,\eta_0)) + \langle \sqrt{W_0}V_o(k,\eta_0) + \sqrt{W_2}V_i(k,\eta_0) \rangle^2,
\end{align}
where $V_o$ and $V_i$ are the unperturbed (`original') and ionospheric components of the expression, respectively, and $\eta_0 = \eta(0)$ is the lowest spatial line-of-sight mode (the DC term). The power spectrum is then,
\begin{equation}
P_{\rm ani} = \frac{1}{\Omega} \langle V(k,\eta)^\ast V(k,\eta)\rangle.
\end{equation}

Note that this model corresponds to a `snapshot' observation of the ionosphere, during which the phase of the ionospheric structures has not evolved. In general, \citet{jordan16} observe anisotropic structures that are persistent in structure (amplitude and wavelength) but evolve in phase as they rotate relative to the celestial sphere (fixed in the geocentric frame, above the MRO site). For extended observations of the same field, this phase evolution will decohere sources, destroying power as the sources change apparent position. This can be quantified by modifying the expected visibility with an integral over an evolving phase term:
\begin{eqnarray}
\langle V(u,v) \rangle &=& \int_{t=0}^T dt \frac{\alpha S_{\rm max}^{2-\beta}}{2-\beta} 2\pi\sigma_{\nu}^2 \exp{(-2\pi^2v^2\sigma_\nu^2)}\, {G}(t),
\end{eqnarray}
with
\begin{eqnarray}
{G}(t) &=& \exp{(-2\pi^2u^2\sigma_\nu^2)} \\\nonumber
&+& \frac{\epsilon(\nu)}{2} \left(\exp{(i\phi(t))}\exp{(-2\pi^2\sigma^2(u-k_l)^2)} - \exp{(-i\phi(t))}\exp{(-2\pi^2\sigma^2(u+k_l)^2)} \right).
\end{eqnarray}
This term acts to decohere the visibility phases, destroying power at the characteristic scale of the structures, with maximal decoherence when the phase evolves by $\pi$ over the coherently-summed observation. For the datasets in question, the phase evolution observed are $\Delta\phi\lesssim 0.5$.

{\bf Full model:}
With these two analytic expressions, Equations \ref{eqn:turb} and \ref{eqn:anis}, for the contributions of the two components of the observed ionospheric activity, the expected impact due to each, and due to a weighted combination of the two, can be explored.

\subsection{Estimating ionospheric parameters from calibration solutions}\label{sec:fim}
As already described, the current MWA EoR algorithms use the calibration and peeling information for the brightest apparent point sources in the field to form an assessment of ionospheric activity in each observation \citep[through the ionospheric metric of][]{jordan16}. They do not then use this information to reconstruct a TEC model to apply to the residual data for correction. Instead, the contaminated data are discarded from the analysis.

However, in principle, these records of apparent source positional shifts can be used to estimate the parameters of an ionospheric model, and apply these to `de-warp' data in the image plane, allowing the observations to be included in EoR analysis. Such algorithms have, and are, being developed for LOFAR \citep[SPAM, ][]{intema_thesis09,prasad14}, MWA \citep{rioja18,hurleywalker16}, and SKA \citep[internal SKA Resolution Teams and][]{rioja18,tasse13,trott15a,cornwell16}. Such an approach has not yet been required by the MWA EoR projects, where the baselines are relatively short, the dataset is large, and only a small fraction were found to contain ionospheric activity. However, for future precision 21~cm science with long baseline arrays (e.g., SKA), this will become an imperative.

As such, we estimate the precision with which the parameters of an ionospheric model could theoretically be estimated with the information available in the MWA calibration solutions, using a Fisher Analysis and the Cramer-Rao Bound \citep[CRB,][]{kay98}. We apply this specifically to the MWA case, but show scalings for a general interferometer.

Our measurement set is the residual visibility set for each baseline, where the model is fitted to the difference between the expected (catalogue) and measured positions of sources in the field:
\begin{equation}
    V_{\rm mod}(\vec{u},;\nu) = V_{\rm meas}(\vec{u};\nu) - V_{\rm catal}(\vec{u};\nu),
\end{equation}
where
\begin{eqnarray}
    V_{\rm meas}(\vec{u};\nu) &=& \displaystyle\sum_{i=0}^{N_p} S_iB_i(\vec{l}+\Delta\vec{l}) \exp{(-2\pi{i}(u(\vec{l}+\Delta\vec{l})))}  \nonumber\\
    &+& \displaystyle\sum_{i=N_p}^{N_T} S_iB_i(\vec{l}+\Delta\vec{l}) \exp{(-2\pi{i}(\vec{u}\cdot(\vec{l}+\Delta\vec{l})))} \\
    V_{\rm catal}(\vec{u};\nu) &=& \displaystyle\sum_{i=0}^{N_p} S_iB_i(\vec{l}) \exp{(-2\pi{i}(\vec{u}\cdot\vec{l}))}.
\end{eqnarray}
Here, the measured data include the $N_p$ sources that are used for calibration, peeled and yield ionospheric model information, and the remaining sky of $N_T-N_p$ unmodelled sources. The latter are fundamentally confused sources, and will be treated as an additional noise term in the Fisher Analysis according to a statistical description of their contribution to visibility noise. Also, $\Delta\vec{l}$ contains the vector apparent positional offsets of each source due to the ionospheric phase screen, and this variable encodes the ionospheric parameters to be fitted:
\begin{equation}
    \Delta\vec{l}(\nu) = \epsilon(\nu) \sin{(2\pi(k_ll+k_lm))} = \epsilon_0 \sin{(2\pi(k_ll+k_lm))} \left( \frac{\nu}{\nu_0} \right)^{-2},
\end{equation}
with $\vec\theta = [\epsilon_0,k_l,k_m]$ as the vector of parameters to be estimated for the CRB in the case of duct-like behaviour. Pre-empting the results of future sections, we restrict our attention to this ionospheric type, because we find it to have impact on the EoR experiments, and because it provides a deterministic model for Fisher Analysis.

The CRB yields the minimum estimation precision for parameter estimation for an unbiased and optimal estimator (which may not exist) where all of the information is used efficiently to estimate model parameters. The Maximum Likelihood Estimator (MLE) is asymptotically efficient (for large datasets and signal-to-noise ratios), and reduces to a weighted least-squares estimator when the data are Gaussian distributed. 
For data that are multivariate generalised Gaussian-distributed, the Fisher Information is encased in a $N_\theta \times N_\theta$ matrix $\mathcal{I}$, where $N_\theta$ is the number of unknown parameters. The $ij$-th element of this matrix is given by:
\begin{equation}
    \mathcal{I}_{ij} = \left( \frac{\partial \vec{V}_{\rm mod}}{\partial \theta_i} \right)^\dagger \mathcal{C}^{-1} \left( \frac{\partial \vec{V}_{\rm mod}}{\partial \theta_j} \right),
\end{equation}
where the sum over the data (baselines) is encoded in the vector model visibility derivatives and $\mathcal{C}$ is the data covariance matrix. This form is appropriate for the case where no unknown parameters are contained within the data covariance model. Before describing the form of the covariance, we extend the expression for the Fisher Information specifically for the current case where the parameters are encoded within the individual source position shifts:
\begin{equation}
    \mathcal{I}_{ij} = \left( \frac{\partial \vec{V}_{\rm mod}}{\partial \Delta\vec{l}} \frac{\partial \Delta\vec{l}}{\partial \theta_i} \right)^\dagger \mathcal{C}^{-1} \left( \frac{\partial \vec{V}_{\rm mod}}{\partial \Delta\vec{l}} \frac{\partial \Delta\vec{l}}{\partial \theta_j} \right).
\end{equation}

The data covariance matrix in its simplest form carries the information on the thermal (radiometric) noise uncertainty on each datum (visibility noise in Jansky indexed by baseline and frequency):
\begin{equation}
    \sigma_{\rm therm} = \frac{2k_BT_{\rm sys}(\nu)}{\sqrt{\Delta\nu\Delta{t}}} \,\, {\rm Jy},
\end{equation}
with $\Delta\nu$ and $\Delta{t}$ describing the spectral and temporal resolution of each measurement, and the system temperature (sky-dominated at these frequencies) described by \citep{briggs06}\footnote{{A more accurate form is \citep{braun13}:
\begin{equation}
    T_{\rm sys}(\nu) \approx 20 (408/\nu_{\rm MHz})^{2.75} + 2.73 + 288 (0.005+0.1314\exp{8(\ln{(\nu_{\rm MHz}/1000)}-\ln{(22.23)})} {\rm K} \nonumber
\end{equation}}}
\begin{equation}
    T_{\rm sys}(\nu) \simeq 180\, {\rm K} \left( \frac{\nu_{\rm MHz}}{180} \right)^{-2.6}.
\end{equation}
In addition to this measurement noise, the estimation of ionospheric parameters occurs within a dataset that contains an unmodelled confused source background. For a single frequency, the point source contribution to the data covariance matrix is uncorrelated between baselines and reduces to (cf, Equation \ref{varsrc}):
\begin{equation}
    {\rm Var}(V(u,v)) = \frac{\alpha S_{\rm max}^{3-\beta}}{3-\beta} \pi\sigma_{\nu}^2 \,\,{\rm Jy}^2,
\end{equation}
with $S_{\rm max}$ corresponding to the peeling limit and $\sigma_\nu$ again representing the primary beam width. This is determined by the dominant image noise term (either the confusion level or radiometric noise). In the case of thermal noise dominance and considering an estimation threshold of F times the noise, the maximum peeling flux density is:
\begin{equation}
    S_{\rm max} = F \frac{\sigma_{\rm therm}}{\sqrt{N_a(N_a-1)/2}} \,\, {\rm Jy}
\end{equation}
for $N_a$ antennas. This can be used to estimate the number of sources that are peeled (and therefore used in the estimation) and their flux density distribution according to a broken power-law parametrisation of the source number density (where the exponent of the flux density steepens from $\beta\simeq{-1.59}$ to $-2.5$ above 1~Jy, yielding fewer bright sources).

The full data covariance therefore reduces to a variance, and each diagonal entry is given by:
\begin{equation}
    \mathcal{C}_{ii} = \sigma^2_{\rm therm} + \frac{\alpha S_{\rm max}^{3-\beta}}{3-\beta} \pi\sigma_{\nu}^2.
\end{equation}

This general model is applied to the specific MWA case. Table \ref{table:params} describes the relevant parameters.
\begin{table}
\centering
\begin{tabular}{|c||c||c|}
\hline
Parameter & Value & Value\\ 
\hline \hline
& 150~MHz & 100~MHz\\
\hline
S$_{\rm max}$ & 200~mJy & 300~mJy \\
N$_p$ & 1200 & 1600\\
N$_a$ & 128 & 128\\
$\Delta{t}$ & 120s & 120s\\
$\Delta\nu$ & 30.72~MHz & 30.72~MHz\\
\hline \hline 
$\frac{\epsilon_0}{\Delta\epsilon_0}$ & 200 & 130 \\
\hline \hline 
\end{tabular}
\caption{Observational parameters for estimating ionospheric parameters for the MWA EoR experiment and output amplitude signal-to-noise ratio.}\label{table:params}
\end{table} 
We implement this model and compute the CRB. Table \ref{table:params} also contains the resulting signal-to-noise ratio on the ionospheric amplitude, $\frac{\epsilon_0}{\Delta\epsilon_0}$. At both frequencies, the large number of sources available to be used to estimation yield high precision estimates. Notably, the precision on the amplitude is factors of hundreds, corresponding to factors of 10$^4$ reduction in power bias. In the predictions for the MWA EoR experiment presented shortly in Section \ref{section:tolerances}, we apply both the current approach (no attempt to correct residual data) and also a theoretical approach where the ionospheric measurements are used to produce an ionospheric model and residual data are corrected before formation of the power spectrum.

\section{Results}\label{section:results}
\subsection{Simulations}
\begin{figure}
{
\includegraphics[width=0.95\textwidth,angle=0]{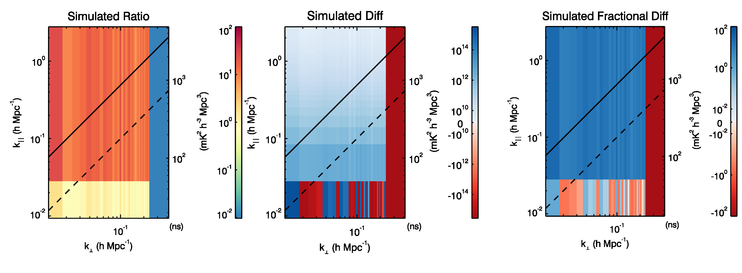}
}
\caption{Simulated prediction for the ratio (active/quiet), difference (active -- quiet) and fractional difference of the anisotropic ionosphere compared with the unperturbed ionosphere. The power is increased on all scales as the coherent structures concentrate the signal in the sky.}
\label{fig:sim_predictducts_act}
\end{figure}

To test the appropriateness of the analytic models we have formed for the two ionospheric components, we perform simple simulations of perturbed point source fields. Figure \ref{fig:sim_predictducts_act} displays the ratio (active divided by quiet), difference (active minus quiet) and fractional difference (difference power divided by unperturbed power) for a simple model with 20,000 point sources distributed across the sky and 40 sky realisations. Anisotropic ionospheric activity is imprinted by altering the probability of a source being located at a physical location, to match a sinusoidal distribution aligned with the $m$-axis (as in Equation \ref{eqn:anis_model}). The characteristic wavelength is $k_l=45$ (1.5 degrees) with a fractional amplitude of 10 percent. The power cutoff above $k_\bot=0.4$ is due to the baseline sampling, not the ionosphere.

\subsection{Analytic predictions}\label{section:analytic_results}
\begin{figure}
{
\includegraphics[width=0.95\textwidth,angle=0]{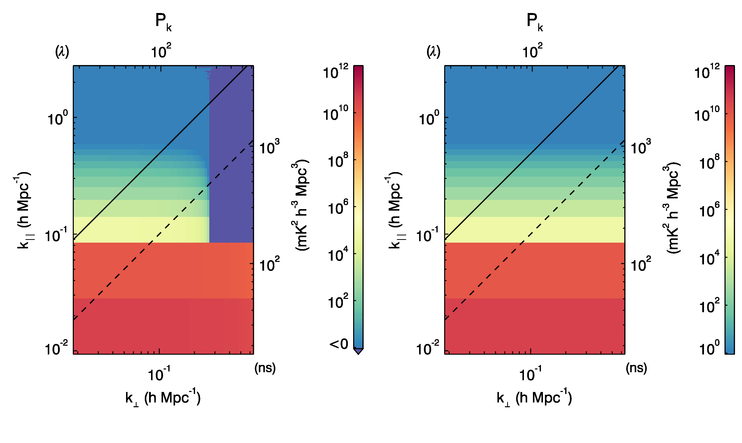}
}
\caption{Analytic prediction for the 2D power spectrum of a turbulent ionosphere (left) and unperturbed ionosphere (right). The turbulent parameters set a scale of $1/u_0=$~2 arcmin and amplitude {$K_0 = 5 \times 10^{-7}$}.}
\label{fig:predict}
\end{figure}

\begin{figure}
{
\includegraphics[width=0.95\textwidth,angle=0]{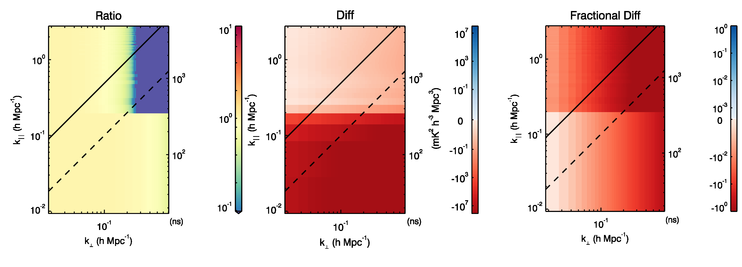}
}
\caption{Analytic prediction for the ratio, difference and fractional difference of the full turbulent ionosphere compared with the unperturbed ionosphere. The primary loss of power occurs on scales smaller than the characteristic turbulent scale.}
\label{fig:sim_predict}
\end{figure}

The analytic predictions for the turbulent and structured components of the observed ionospheric activity can be compared with the unperturbed models. The turbulent component power spectra are displayed in Figures \ref{fig:predict}-\ref{fig:sim_predict}, showing the perturbed (left) and unperturbed (right) spectra for a characteristic scale of $1/u_0$=~2 arcmin, and amplitude, {$K_0 = 5 \times 10^{-7}$}, and their ratio and difference. This corresponds to an effective power cut-off at $1/u_{\rm cut}\sim$4 arcmin, which is larger than the characteristic scale observed in the data ($1/u_{\rm cut}\sim$0.2 arcmin), but allows us to explore how the turbulence manifests in the power spectrum.

\begin{figure}
{
\includegraphics[width=0.95\textwidth,angle=0]{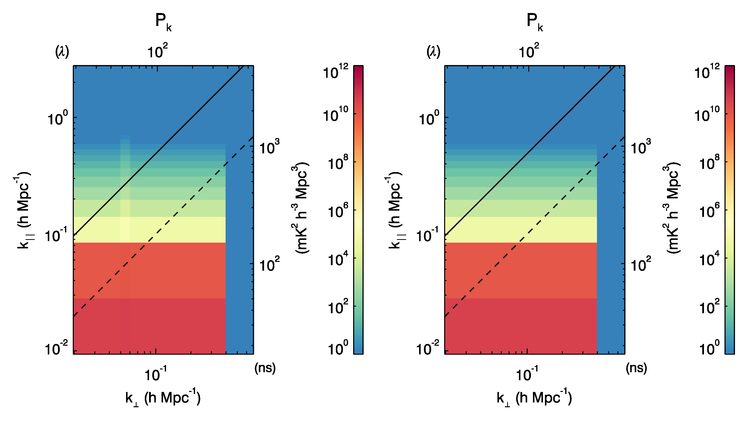}
}
\caption{Analytic prediction for the power spectrum of a snapshot anisotropic ionospheric perturbation, characterised by coherent density variability of the foreground sources (left), compared with the unperturbed prediction (right). The characteristic scale is $\sim$1.5~degrees, with a 10\% amplitude.}
\label{fig:predictducts}
\end{figure}

\begin{figure}
{
\includegraphics[width=0.95\textwidth,angle=0]{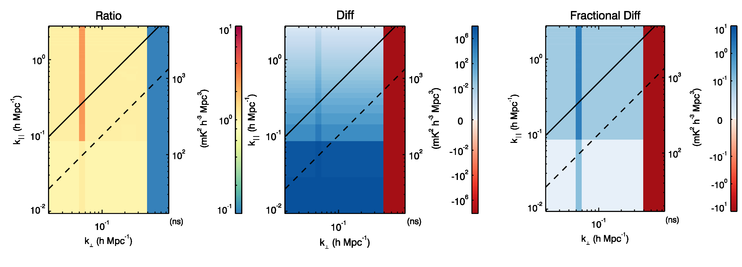}
}
\caption{Analytic prediction for the ratio, difference and fractional difference of the anisotropic ionosphere compared with the unperturbed ionosphere. The power is increased on all scales as the coherent structures concentrate the signal in the sky.}
\label{fig:sim_predictducts}
\end{figure}

The snapshot anisotropic component is similarly shown in Figures \ref{fig:predictducts} and \ref{fig:sim_predictducts} with $k_l$=45, corresponding to a scale of $\sim$1.5~degrees, and a 10\% amplitude.

There are key differences between these two modes of activity. While the turbulent component is shown to destroy power on all scales, it is primarily on scales smaller than the characteristic turbulent scale that are most affected. The fractional power loss on larger scales is very small and mostly corresponds to the differing spectral behaviour. When considering data with thermal noise, the turbulent component is expected to impact only at smaller scales. The anisotropic mode, however, displays increased power, because the ducts are effectively concentrating signal on the sky, leading to less decoherence between (physically unassociated) foreground radio galaxies. This concentration occurs only in one dimension, spreading the increased power across all angular modes in the average of $k_\bot = \sqrt{u^2+v^2}$.

\subsection{Comparison with data}
\begin{figure}
{
\includegraphics[width=0.95\textwidth,angle=0]{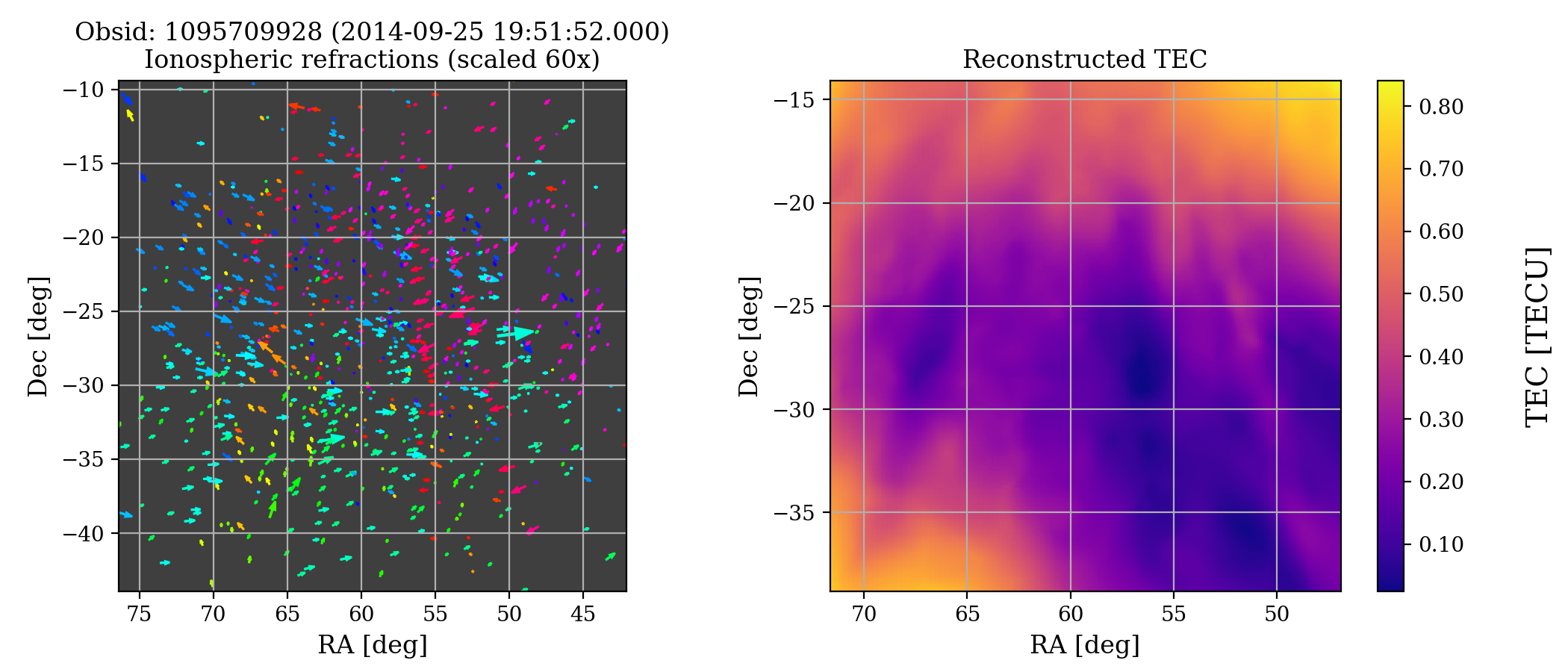}
}
\caption{Example reconstruction of slant TEC field from the measured point source deviations from catalogue positions, for a two-minute observation from 2014 September 25, displaying anisotropic duct-like behaviour. (Left) Vectors of point source positional offsets; (right) Reconstructed slant TEC scalar field. \citep{jordan16}.}
\label{fig:recon}
\end{figure}

\begin{figure}
{
\includegraphics[width=0.95\textwidth,angle=0]{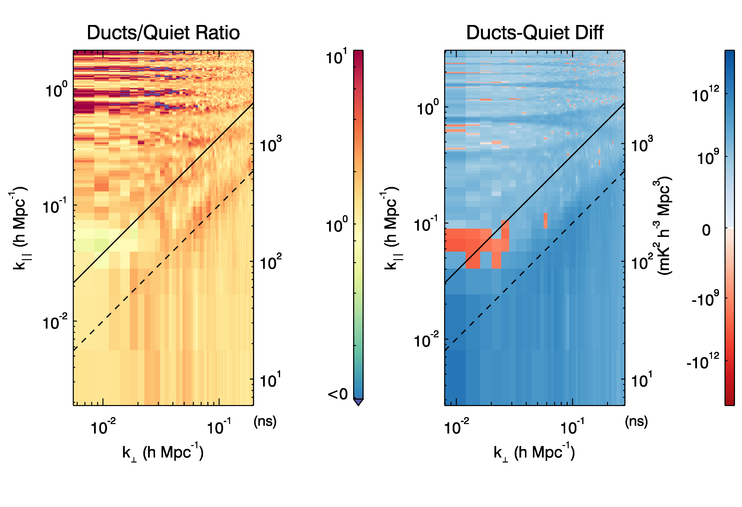}
}
\caption{Ratio and difference of 30 minutes of zenith-pointed observations from September 2014, during which the MWA observed Type IV (anisotropic structures) and Type I (quiet) conditions. Each dataset is formed from data that are contiguous in time, with the same observation parameters, phase centre and a zenith-pointed beam.}
\label{fig:data}
\end{figure}

\begin{figure}
{
\includegraphics[width=0.95\textwidth,angle=0]{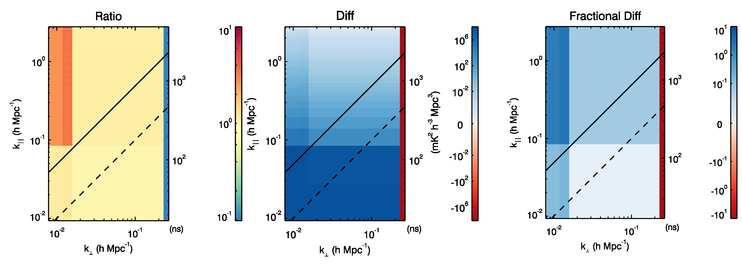}
}
\caption{Analytic prediction for the ratio, difference and fractional difference of the anisotropic ionosphere compared with the unperturbed ionosphere for structured ionospheric parameters consistent with that observed in the 2014 data.}
\label{fig:datacomp}
\end{figure}

Figure \ref{fig:recon} displays data from a zenith-pointed observation from 2014 September 25 of the EoR1 field (RA=4h, Dec.=$-$30 degrees) corresponding to anisotropic duct-like behaviour. The apparent source offsets are used to reconstruct a slant TEC field across the FOV, from which an ionospheric metric of activity can be derived and a spatial power spectrum formed.
In these 30 mins of contiguous data, the coherent structures are seen to move with respect to the geocentric frame, but mostly maintain their position in the celestial frame, consistent with a slowly moving local ionospheric structure above the observatory. The visibilities are formed with respect to a phase centre that is fixed in the celestial frame, corresponding to an ionospheric phase screen that is constant with time for these data. Therefore we would expect minimal decoherence of visibilities that are coherently combined over the 30 mins. The observed wavelength of the structures is $\sim$5--8 degrees, corresponding to $k_l\sim 10$. Figure \ref{fig:data} shows the ratio and difference for these data, compared with a matching set from the same week with quiet conditions.
A comparison analytic prediction for these parameters is shown in Figure \ref{fig:datacomp}.
The prediction is broadly consistent with the data: the overall power is increased, with the primary change occurring at large scales, consistent with the duct structures observed in the reconstructed scalar TEC fields (see Figure \ref{fig:recon}).

\section{Tolerances}
\label{section:tolerances}
Despite individual Types of ionospheric conditions being observed at the MRO, it is the integrated effect of these that will decohere otherwise coherent visibility data, and destroy power on scales corresponding to that for the ionospheric activity. \citet{jordan16} characterised four qualitatively-distinct types of activity, and their relative occupancy of EoR data from the 2015 observing season, but each Type will affect the power spectrum in a different way. For the Type II (turbulent) mode, occupying 15\% of that dataset, the typical median positional offset is $\sim$0.2 arcmin, corresponding to a scale of little relevance to EoR science ($k_\bot=16 h$~Mpc$^{-1}$). Because the turbulent mode primarily destroys power on scales smaller than the characteristic scale, this mode is not likely to be relevant for the EoR experiment\footnote{A separate question, not addressed here, is the ability of the direction-dependent calibration and peeling of bright sources to correctly measure this and remove power cleanly. This is likely to be a much larger effect than residual power loss from turbulence shifting the positions of weaker sources.}. The anisotropic modes, however, routinely display activity on degree scales, and these require further investigation into their impact.

The analytic predictions for bias in the power spectrum space must be compared with expectations for cosmological signal strength for a given experiment, redshift range, and spatial scale. This allows tolerances to be set for the degree of ionospheric activity acceptable in the data before it is predicted to bias the experimental result. This can be applied to both the uncorrected and corrected datasets.

To estimate these tolerances, we used simulated cosmological power spectra as a function of spatial scale, and at different redshifts, from 21cmFAST \citep{mesinger11} and assuming the favoured faint galaxy model for reionisation. These simulated power spectra are used as reference EoR signals. Power spectra are extracted for redshifts, $z=8.5,13.2$, and we employ the spherically-averaged dimensionless power spectrum:
\begin{equation}
\Delta^2(k) = \frac{P(k)k^3}{2\pi^2} {\rm mK}^2.    
\end{equation}

Given that the ionosphere is not of a persistent structure over the full course of a $\sim$1000-hour experiment, the timescale of power bias is relevant. Observations using the MWA at the MRO in EoR data, reported by \citet{jordan16}, demonstrate that each 4-hour observing night typically shows persistent structure (with a few notable exceptions where activity changes rapidly due to external factors). Therefore, we assume that the ionospheric activity coherence time is equivalent to this timescale. We can therefore use the observed occupancy of different ionospheric Types at the MRO to assess the impact on the EoR power spectrum. The instrument model considers the station size and array layout for the MWA Phase I (relevant for comparing with 2014 data), to compute the relevant foreground signal in the data, and the resolution of the $uv$-plane.

\begin{figure}
{
\includegraphics[width=0.85\textwidth,angle=0]{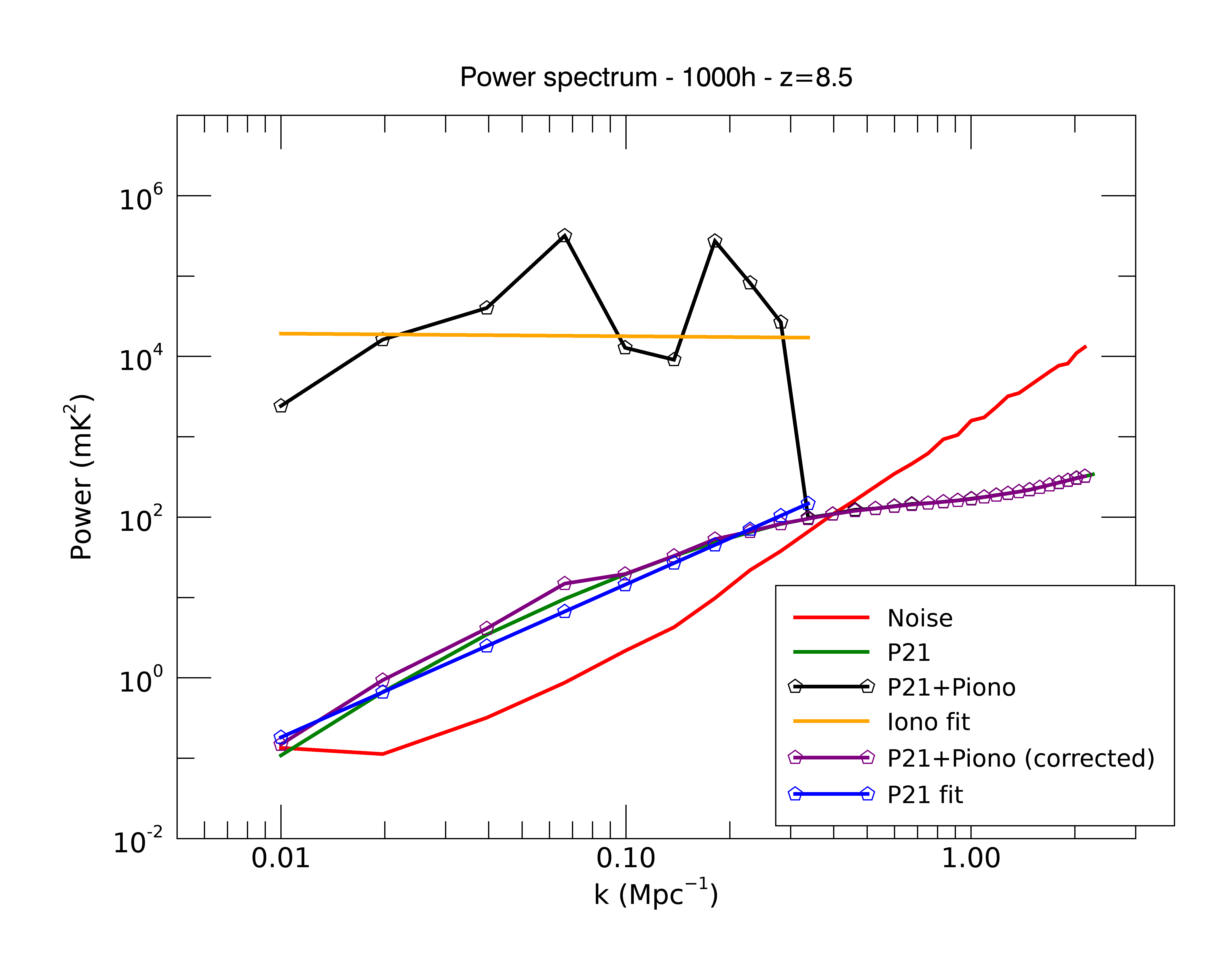}
}
\caption{Spherically-averaged dimensionless power spectrum for a simulated 21cm EoR signal at $z=8.5$ (150~MHz), when no ionospherically-active data are included (green) and when 8\% of data contain anisotropic activity (black=uncorrected, purple=corrected). Linear fits to the logarithmic slope in SNR$>$1 regions are also displayed (orange, blue).}
\label{fig:mwa_z8pt5}
\end{figure}

\begin{figure}
{
\includegraphics[width=0.85\textwidth,angle=0]{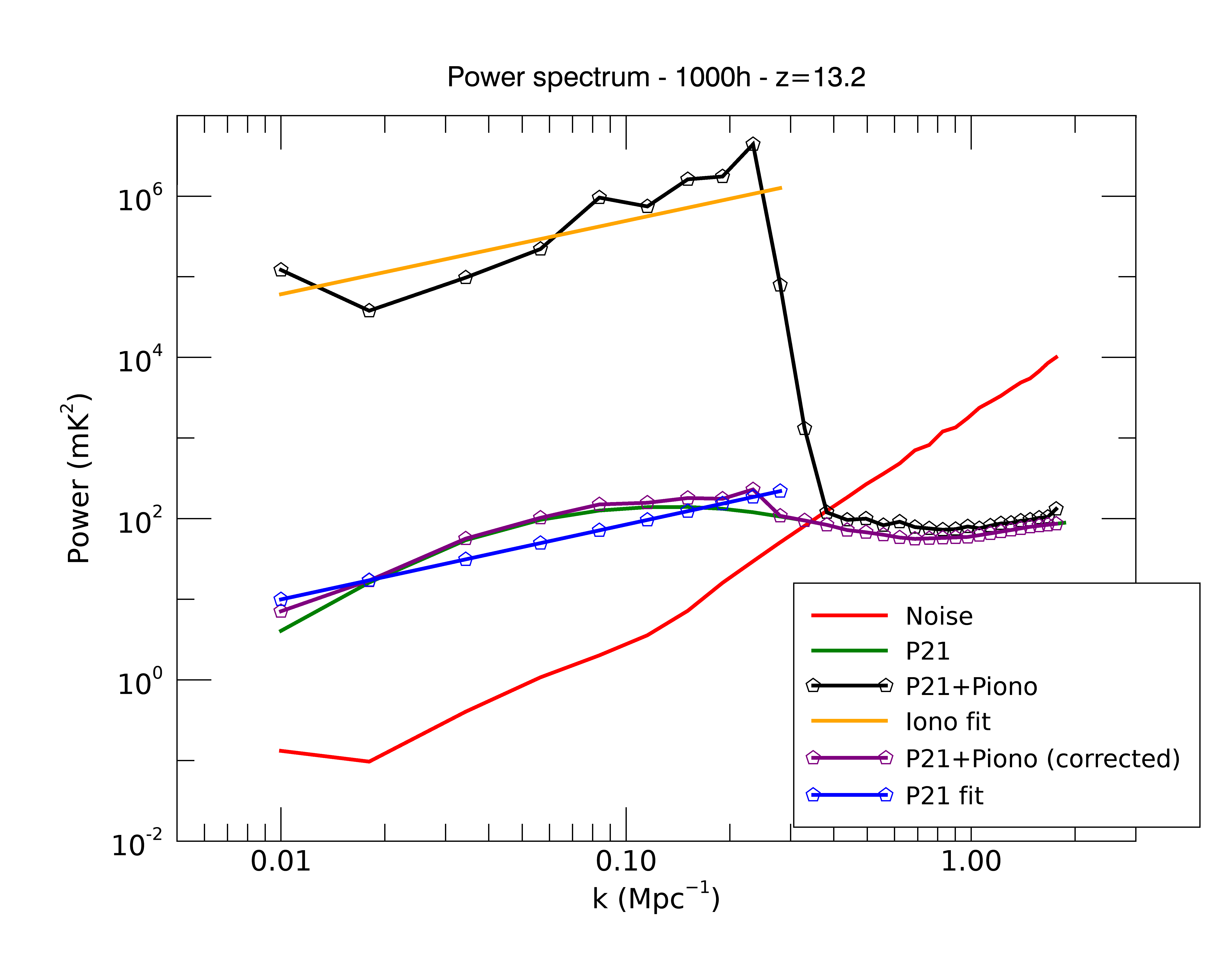}
}
\caption{Spherically-averaged dimensionless power spectrum for a simulated 21cm EoR signal at $z=13.2$ (100~MHz), when no ionospherically-active data are included (green) and when 8\% of data contain anisotropic activity (black=uncorrected, purple=corrected). Linear fits to the logarithmic slope in SNR$>$1 regions are also displayed (orange, blue).}
\label{fig:mwa_z13pt2}
\end{figure}

We start by considering the uncorrected residual signal model, as is currently implemented in the MWA EoR pipelines.
The analytic model described in Sections \ref{section:analytic_model} and \ref{section:analytic_results} is used to compute the impact of ionospheric for the MWA experiments at relevant redshifts ($z=8.5, 13.2$ corresponding to 150~MHz and 100~MHz, respectively). As a metric for the realistic impact of including active data, we compute the bias in the slope of the 1D 21cm power spectrum when residual power from activity is included. As described above, reference EoR signal power is extracted from simulated models for reionisation \citep{mesinger11}. We consider the relative observed fractions of each Type of ionospheric activity at the MRO to construct the model observed power spectrum if all data were included. E.g., allowing 8\% of the data to contain anisotropic structures and 15\% to contain turbulent modes. We consider a 1000~h experiment with a bandwidth of 10~MHz. Figures \ref{fig:mwa_z8pt5} and \ref{fig:mwa_z13pt2} display the cosmological signal alone (green), cosmological signal plus ionospheric foreground bias (black), the thermal noise level (red), and linear fits to the logarithmic slope in the region where the signal-to-noise ratio exceeds unity (orange, blue). The structure in the biased power is due to the cylindrical averaging over modes with differing levels of foreground contamination. Omitting these modes (`working outside of the wedge') leaves few $k$-modes available with SNR$>$1.

Also shown in these figures is the corrected case (purple), where we have propagated the uncertainties on the ionospheric amplitude and angular scale factors computed in Section \ref{sec:fim} to the residual (post data correction) power spectrum. The correction here is marginally adequate to be able to use these data for the EoR experiments.

At both redshifts, the ionospheric bias from including 8\% of data from ionospherically-active nights (assuming the foregrounds are not treated with knowledge of the activity and no correction is applied) leads to systematic biases in the 21~cm parameter estimation. Note that both of these cases show extreme biases, implying that there are large fractions of $k$-space where the ionosphere makes the experiment impossible. However, these are for the extreme case where almost 10\% of the data display coherent, structured behaviour, and where no point sources have been correctly extracted. For a less extreme case where perhaps only one percent of the data contains structure, and these data are included in the analysis, the bias will be a factor of 100 lower. In addition, if the brightest sources are well-measured and subtracted cleanly, then the bias will again be reduced. Nonetheless, the potential for a large power bias implies that exclusion of data with active ionospheres is prudent for such a precision experiment.

\begin{figure}
{
\includegraphics[width=0.95\textwidth,angle=0]{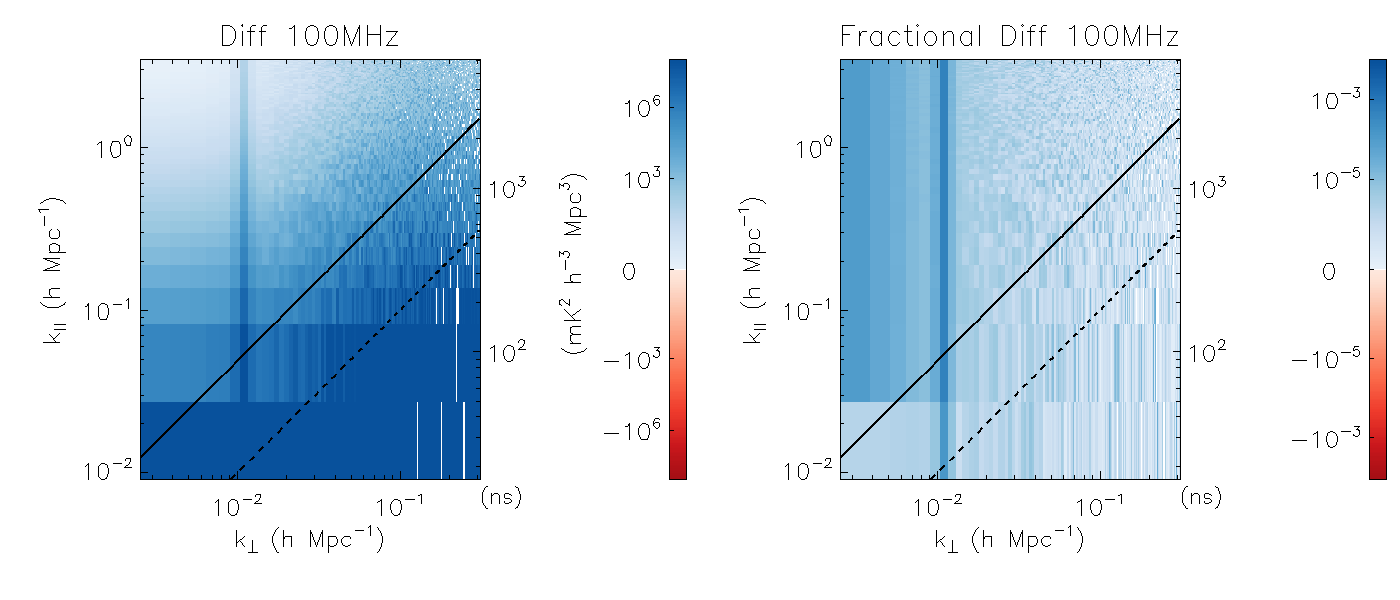}
}
\caption{Difference (left) and fractional difference (right) for an MWA experiment at 100~MHz ($z=13.2$), with anisotropic turbulence occurring at the amplitude observed in 2014 data.}
\label{fig:mwa_z13pt2_diff}
\end{figure}

Figure \ref{fig:mwa_z13pt2_diff} further displays the difference and fractional difference for the uncorrected case at $z=13.2$. Here the instrumental sampling has been included and the foreground wedge is present. Although there is a maximum of 10$^6$ mK$^2$ Mpc$^3$ difference in the 2D power spectrum, when averaged to 1D it corresponds to a factor of 10$^4$ mK$^2$ increase on EoR scales.

We therefore conclude that inclusion of ionospherically-active data (where no ionospheric correction has been applied) will lead to systematic biases in the outputs of MWA EoR experiments.

{When the information is used to estimate the ionospheric model parameters with an ideal estimator, and these are then used to correct the residual data, the bias is reduced by a factor of $\sim$10$^4$, yielding acceptable levels for EoR experiments with the MWA.

\section{Discussion and conclusions}
We have presented analysis of the bias in the EoR power spectrum due to unaccounted residual point source foreground contribution when ionospheric activity is present. This analysis pertains to the case where the activity is assessed, but residual data (after peeling of bright sources) is not corrected and retains the imprint of the activity, and the case where the residual data are corrected with a model formed from the peeling information. In the former case, aside from the bias in the power, combining active and inactive data coherently (averaging visibilities) will lead to decoherence of the signal and will have smaller effects on the 21~cm power spectrum.

Of the two primary Types of ionospheric activity identified at the Murchison Radioastronomy Observatory in MWA EoR datasets (Type I: isotropic small-scale turbulence, and Type IV: turbulent, anisotropic structure) and in other MWA datasets \citep{loi16,hurleywalker16,rioja18}, the scale of the turbulence is such that it should not affect EoR power spectra (where power is principally on larger scales), while anisotropic behaviour (observed to occur in 8\% of observations) imprints residual foreground power that can affect parameter estimation. We therefore conclude that data with anisotropic ionospheric activity should be omitted from deep EoR observations, and data showing turbulent behaviour should be monitored to ensure the spatial scales are not of relevance to the EoR.

In future, with the construction of the low-frequency component of the SKA at the MRO site, expected ionospheric activity can be guided by these results. We expect the same distribution of ionospheric classes (on average) to be present for the SKA-Low experiments. For the SKA, however, the much longer baselines demand ionospheric correction to be performed. While the MWA's short baselines allow for a single phase screen model to be fitted across the array, longer baselines demand more sophisticated models, where individual sources are observed through different ionospheric conditions across the array. The array has been designed to allow for enough ionospheric pierce points to be available to calibrate the instrument, particularly at the challenging lower frequency of 50~MHz. The HERA array, in South Africa, as the other upcoming EoR array, shares a similar compact configuration as the MWA, and its ionospheric calibration requirements are likely to be similar to the MWA, and less stringent than SKA-Low.

\section*{Acknowledgements}
We thank the anonymous referee for a careful reading of the paper, useful comments, and correction of errors, which have all improved the manuscript substantially.
This research was supported by the Australian Research Council Centre of Excellence for All Sky Astrophysics in 3 Dimensions (ASTRO 3D), through project number CE170100013. The International Centre for Radio Astronomy Research (ICRAR) is a Joint Venture of Curtin University and The University of Western Australia, funded by the Western Australian State government. This scientific work makes use of the Murchison Radio-astronomy Observatory, operated by CSIRO. We acknowledge the Wajarri Yamatji people as the traditional owners of the Observatory site. Support for the operation of the MWA is provided by the Australian Government (NCRIS), under a contract to Curtin University administered by Astronomy Australia Limited. We acknowledge the Pawsey Supercomputing Centre which is supported by the Western Australian and Australian Governments.
CMT is supported by an ARC Future Fellowship under grant FT180100196.


\section*{Appendix: Turbulence power spectrum model}
While the effect of high-amplitude features of the ionosphere such as TIDs may be individually modeled, it is expected that a temporally dynamic low-amplitude statistical field exists at all times.
The form of this field is further expected to exhibit Kolmogorov-like turbulence, i.e. an isotropic angular power spectrum with power-law dependence on angular modes $u$, with slope $\kappa \sim -5/3$. In this appendix, we describe the statistical effect of this turbulent field on the observed EoR + foreground signal.

We will exclusively work in the regime in which the flat-sky approximation holds, i.e. in which angular scale $u$ and angle $l$ are related as Fourier duals. 
We furthermore take the Fourier convention most popular in radio interferometry, i.e. $\hat{f} = \int f \exp(-iu\cdot l)$.

We denote the (scalar) TEC field by $\psi(\vec{l})$ and note that in the weak-field regime with which we are here concerned, the first-order effect of the ionosphere is to impose a shift on the observed angular position of incoming sight-lines. 
This shift can be written
\begin{equation}
\label{eq:tec_shift}
    \Delta \vec{l} = K(\nu) \vec{\nabla} \psi(\vec{l}),
\end{equation}
where $K \propto \nu^{-2}$ is a scaling constant.

Our problem bears immediate similarities to that of weak lensing of the CMB.
In both cases, a background angular signal is shifted by the gradient of a scalar field.
Thus, to solve our problem we follow the derivation of \cite{Lewis2008} (see especially \S 4.1). 
Briefly, the observed field (we shall denote the field $S$ to signify flux-density, though it is interchangeable for temperature) is defined by
\begin{equation}
  \tilde{S}(\vec{l}) = S(\vec{l} + \Delta\vec{l}).
\end{equation}
The RHS here affords a series expansion which is a good approximation on our scales of interest for a weak $\psi$. 
Truncating the series at quadratic order and taking the Fourier Transform, while maintaining independence of the $S$ and $\psi$ fields, yields an observed power spectrum \citep[cf.][Eq. 4.12]{Lewis2008}
\begin{equation}
    \label{eq:power_observedapp}
  {P}_{\rm turb} \approx (1 - K(\nu) u^2 R^\psi) P_u^S + K(\nu) P_u^\psi \star P_u^S,
\end{equation}
with $\star$ indicating convolution, and
\begin{equation}
R^\psi = \pi \int u^3 P_u^\psi du.
\end{equation}

The final term in Eq. \ref{eq:power_observedapp} is sub-dominant due to the weak TEC field. Indeed, when the field is strong and the convolution becomes important, the approximations made here themselves become invalid. We thus ignore the convolution for the remainder of this analysis.

The dominant effect of the turbulent ionosphere then is to destroy small-scale power. 
Specifically, the observed power falls to zero at a scale
\begin{equation}
    u' = \sqrt{\frac{1}{K(\nu)R^\psi}}.
\end{equation}
The cut-off is rather sharp (in log-space), and we propose that the most effective way to treat the ionospheric distortion is as a sharp cut in EoR window at some $k_{\rm max}$. 

We suppose that the power spectrum of $\psi$ is Kolmogorov-like, but has some scale above which the power is truncated. In practice, setting this scale above the resolution limit of an instrument serves to eliminate sensitivity to this choice. 
Thus we consider 
\begin{equation} 
P^\psi_u = \begin{cases} \left(\frac{u}{u_0}\right)^\kappa & u \leq u_{\rm max} \\ 0 & u>u_{\rm max}. \end{cases}
\end{equation}
In this case, we have 
\begin{equation}
R^\psi = \frac{u_{\rm max}^{4+\kappa}}{(4+\kappa)u_0^\kappa}.
\end{equation}

Suppose we wish to place a hard cut on the spectrum at $u_f$, when the power is diminished by a fraction $f$ of its original amplitude. 
Further suppose that we set $u_{\rm max}$ to be the limiting resolution of our telescope at a given frequency. 
The natural question to ask then is how the scale $u_f$ compares to $u_{\rm max}$, by considering their ratio. If it is greater than unity, then we needn't worry about the effects of the ionosphere, whereas if the opposite is true, then the ionosphere begins to encroach on the observed modes.
This scale can be written
\begin{equation}
 u_f = \sqrt{\frac{f(4+\kappa)}{K(\nu)}  \left(\frac{u_0}{u_{\rm max}}\right)^\kappa  } \frac{1}{u_{\rm max}^3}.
\end{equation}



\bibliographystyle{aasjournal}
\bibliography{pubs} 



\end{document}